\documentclass[12pt]{article}

         \usepackage{amsfonts}
         \usepackage{amssymb}

\usepackage{graphicx}

\def\be{\begin{eqnarray}}
\def\ee{\end{eqnarray}}
\def\bee{\begin{eqnarray*}}
\def\eee{\end{eqnarray*}}

\newcommand{\proj}[1]{|{#1} \kb {#1}|}

\newtheorem{thm}{Theorem}

\newtheorem{lemma}[thm]{Lemma}

\newtheorem{defn}[thm]{Definition}
         \def\pf{\medbreak\noindent{\bf Proof:}\enspace}
      \def\rmk{\medbreak\noindent{\bf Remark:}\enspace}
         
         \def\qed{{\bf QED}}

       \def\imp{\Rightarrow}
              \def\half{{\textstyle \frac{1}{2}}}
         \def\tr{{\rm  Tr}}

\def\wh{\widehat}
\def\wt{\widetilde}
\def\ds{\displaystyle}
  \def\ts{\textstyle}
\def\bra{\langle}
\def\ket{\rangle}
\def\kb{ \ket \bra }
\def\dg{\dagger}
\def\ot{\otimes}

\def\half{{\textstyle \frac{1}{2}}}

\def\bT{{\mathbf{\bold T}}}

\def\b0{{\mathbf{\bold 0}}}

\def\nl{\newline}

  \newcommand{\braket}[2]{\langle #1 | #2 \rangle}

           \title{Entanglement Breaking Channels}

         \author{Michael Horodecki \\ Institute of Theoretical
Physics and Astrophysics \\ University of Gda\'nsk,
80-952 Gda\'nsk, Poland \\
{\normalsize fizmh@univ.gda.pl } \and Peter W. Shor \\
AT \&T Labs Research \\ Florham Park, New Jersey 07922 USA
\\ {\normalsize shor@research.att.com} \and
Mary Beth Ruskai\\ Department of Mathematics, Tufts University\\
   Medford, Massachusetts 02155 USA \\ {\normalsize ruskai@attbi.com}}

\begin{document}

\maketitle

\begin{abstract}
This paper studies the class of stochastic maps, or channels,
for which $(I \ot \Phi)(\Gamma)$ is always separable
(even for entangled $\Gamma$).  Such maps are called entanglement
breaking, and can always be written in the form
$\Phi(\rho) = \sum_k R_k ~ \tr \, F_k \rho$
where each $R_k$ is a density matrix and $F_k > 0$.
If, in addition, $\Phi$ is trace-preserving,
the  $\{F_k\}$ must form
a positive operator valued measure (POVM).
Some special classes of these maps are considered
and other characterizations given.

Since the set of entanglement-breaking trace-preserving
maps is convex, it can be characterized by its extreme points.
The only extreme points of the set of completely
positive trace preserving maps which are also entanglement breaking
are those known as classical quantum or CQ.  However, for $d \geq 3$,
the set of entanglement breaking maps  has additional extreme points
which are not extreme CQ maps.

\end{abstract}



\section{Introduction}

A quantum channel is represented by a stochastic map,
i.e., a map which is both completely positive and  trace-preserving.
We will refer to these as  CPT  maps.
In this paper we consider the special class of
quantum channels which can be simulated by a classical
channel in the following sense:  The sender makes a
measurement on the input state $\rho$, and send the
outcome $k$ via a classical channel to the receiver who
then prepares an agreed upon state $R_k$.
Such channels can be written in the  form
\be \label{eq:holv}
   \Phi(\rho) = \sum_k R_k \, \tr  F_k \rho
\ee
where each $R_k$ is a density matrix and the  $\{F_k\}$ form
a positive operator valued measure POVM.   We call this
the ``Holevo form'' because it
was introduced by Holevo in \cite{Hv}.

It is also natural consider the class of channels which
break entanglement.
\begin{defn}  \label{def:entbk}
A stochastic map  $\Phi$ is called  {\em entanglement breaking}
if \linebreak $(I \ot \Phi)(\Gamma)$ is always separable, i.e.,
any entangled density matrix  $\Gamma$ is mapped to a separable one.
\end{defn}
It is not hard to see that, as shown in the next section,
a map is entanglement-breaking if and only if it can be
written in the form
\be \label{eq:rank1}
   \Phi(\rho) = \sum_k
      |\psi_k \kb \psi_k| \bra \phi_k, \rho \, \phi_k  \ket
\ee
in which case it is necessarily completely positive.
Furthermore, $\Phi$ is trace-preserving if and only if
$\sum_k | \phi_k \kb \phi_k | = I$, in which case,
(\ref{eq:rank1}) is a special case of (\ref{eq:holv}).
One can show that the converse also holds, so that we
have the following result.
\begin{thm}  \label{thm:holv.iff.entbk}
A channel  can be written in the  form {\em (\ref{eq:holv})}
using positive semi-definite operators $F_k$
if and only if it is entanglement breaking.  Such a map
is also trace-preserving if and only if the $\{F_k\}$ form
a POVM or, equivalently, $\sum_k | \phi_k \kb \phi_k | = I$.
\end{thm}
The rather straightforward proof will be given in the next
section together with some additional equivalences.
We will refer to stochastic maps which are both entanglement-breaking
and   trace-preserving as EBT.

Of course there are stochastic maps which are not of the
form (\ref{eq:holv}).  In particular, conjugation with a unitary
matrix is not EBT.
Channels which break entanglement are particularly noisy
in some sense,  e.g., a qubit map is EBT
if the image of the Bloch
sphere collapses to a plane or a line.  In the opposite
direction, we will show
that a channel in $d$ dimensions is {\em not} EBT
if it can be written using fewer than $d$ Kraus operators.

\begin{thm}  \label{thm:entbk.conv}
The set of EBT maps is convex.
\end{thm}
Although this follows easily from the definition of  entanglement breaking,
it may be  instructive to also show directly that the set of maps of the
form (\ref{eq:holv}) is convex.  Let $\Phi$ and
$\widetilde{\Phi}$ denote such maps with density matrices
$\{ R_j \}_{j = 1 \ldots m}$ and
$\{ \widetilde{R}_k \}_{ k = 1 \ldots n}$  and POVM's
$\{ E_j \}_{j = 1 \ldots m}$ and
$\{ \widetilde{E}_k \}_{ k = 1 \ldots n}$ respectively.
For any $\alpha \in [0,1]$ the map
\bee
   \big[ \alpha \Phi + (1- \alpha ) \widetilde{\Phi} \big] (\rho) =
      \sum_j R_j \, \tr ( \alpha E_j \rho) +
    \sum_k \widetilde{R}_k \, \tr [ (1- \alpha)\tilde{E}_j \rho]
\eee
has the  form (\ref{eq:holv})  since
$\{ \alpha E_1, \alpha E_2 , \ldots  \alpha E_m,
    (1- \alpha)\widetilde{E}_1, \ldots (1- \alpha)\widetilde{E}_n \}$ is
also a POVM.

Note that we have used implicitly the idea of generating
a new POVM as the convex combination of two POVM's,   In this
sense, the set of POVM's is also convex, and one might expect
that the extreme points of the set of entanglement-breaking
maps are precisely
those with an extreme POVM and pure $R_k$.  However, this is false; at
end of Section~3 of \cite{EBQ}, the trine POVM is
used to give an example of a qubit channel
which is not extreme, despite the fact that the POVM is.

Certain subclasses of EBT maps
  are particularly important.  Holevo called a channel
\begin{itemize}
\item  {\em  classical-quantum} (CQ) if each
$F_k = | k \kb k |$ in the POVM is a one-dimensional projection.
In this case, (\ref{eq:holv})  reduces to
$ \Phi(\rho) = \sum_k R_k ~  \bra k, \rho k \, \ket$.
\item  {\em  quantum-classical} (QC) if each
density matrix $R_k = | k \kb k |$ is a one-dimensional projection and
$\sum_k R_k = I$.
\end{itemize}
If a CQ map has the property that each density matrix
$R_k = | \psi_k \kb \psi_k|$ is a pure state,
we will call it an {\em extreme CQ} map.
  Note that the
pure states $| \psi_k \ket$ need not be orthonormal, or even
linearly independent.   We will see in Section 3 that
extreme CQ maps are always extreme points of the set of EBT maps,
but they are only extreme points for the set of CPT maps if
   all pairs $\bra \psi_j , \psi_k \ket$ are non-zero.

When all $R_k = R$ are
identical, then $\Phi$ is the maximally noisy map
  $\Phi(\rho) = R $ for all $\rho$.
Because it maps all density matrices to the same
$R$, its image is a single ``point'' in the set of density matrices
and its capacity is zero.  A point channel is extreme if and
only if its image $R$ is a pure state.   A {\em point}
channel is a special case of a CQ map; however, because all $R_k = R$ the
sum in (\ref{eq:holv}) can be reduced to a single term with $E_1 = I$.
For $d > 2$, one can also consider those CQ maps for which some $R_k$
are identical; then
the POVM can be written as a projective measurement, and the
image is a polyhedron.

It is useful to have Kraus operator representations of EBT maps.
For $\Phi$ of the form (\ref{eq:holv}), let
$A_{kmn} = \sqrt{R_k}\,  |m \kb n | \, \sqrt{F_k}$
where $\{ |m \ket \}$ and $\{ |n \ket \}$ are orthonormal bases.
Then one easily verifies that
\be \label{eq:Kraus.d}
     \sum_{kmn} A_{kmn} \, \rho \, A_{kmn}^{\dg} = \sum_k R_k \tr F_k \rho
\ee
For CQ and QC maps these operators reduce to
$A_{km} = \sqrt{R_k}\,  |m \kb k |$ and
$A_{kn} =   |k \kb n | \, \sqrt{F_k}$ respectively.
Moreover, if all density matrices are pure states
$R_k = |\psi_k \kb \psi_k|$, then one can achieve a further
reduction to $A_k = | \psi_k \kb k |$ in the case of CQ maps.


Holevo \cite{Hv} showed that for EBT maps
the  Holevo  capacity  (i.e., the capacity of a quantum channel
used for classical communication with product
inputs) is additive.    This result
was  extended by King \cite{King1} to additivity of
the capacity of channels of the form $\Phi \ot \Omega$ where
$\Phi$ is CQ or QC and $\Omega$ is completely arbitrary.
Shor \cite{Shor} then proved   the additivity of
minimal entropy and Holevo capacity when $\Phi$ is EBT
and $\Omega$ arbitrary.   Quite recently, King \cite{King2}
showed that the maximal p-norms of EBT channels are multiplicative,
and used this to give another proof of Shor's additivity results for
minimal entropy and Holevo capacity.   In a related
development, Vidal, D\"{u}r and Cirac \cite{VDC} used
Shor's techniques to prove additivity of the entanglement
of formation for a class of mixed states associated with
EBT maps.

Because
it is important to understand the differences between those
channels which break entanglement and those which preserve it,
we seek other characterizations of these channels, describe
their extreme points,  and  examine their properties.
Results for qubits are
given in a related paper \cite{EBQ} which follows.
Some analysis of entanglement breaking  channels was
also independently presented by
Verstraete and Verschelde~ \cite{VV}.



\section{Equivalent conditions}

In this section, we establish a number of equivalent
characterizations of EBT maps, some of which were already
discussed in the previous section.

\begin{thm}  \label{thm:gen}  The following are equivalent
\begin{itemize}
\item[A)] $\Phi$ has the Holevo form (\ref{eq:holv}) with
$F_k$ positive semi-definite.
\item[B)] $\Phi$ is entanglement breaking.
\item[C)] $(I \ot \Phi)( | \beta \kb \beta | )$ is separable
for $|\beta \ket = d^{-1/2} \sum_j |j \ket \ot |j \ket $
a maximally entangled  state.
\item[D)] $\Phi$ can be written in operator sum
form using only Kraus operators of rank one.
\item[E)] $\Upsilon  \circ  \Phi$ is completely positive for
all positivity preserving maps $\Upsilon$.
\item[F)] $\Phi \circ \Upsilon $ is completely positive for
all positivity preserving maps $\Upsilon$.
\end{itemize}
A corresponding equivalence holds for CPT and EBT maps with the
additional conditions that $\{F_k \}$ is a POVM, the Kraus
operators $A_k$ satisfy $\sum_k A_k^{\dg} A_k = I$, and $\Upsilon$
is trace-preserving.
\end{thm}

To  prove this result, we will make use of the
correspondence \cite{Choi,J} between maps and states
given by $\Phi \leftrightarrow (I \ot \Phi)( | \beta \kb \beta | )$
(Also see \cite{BDSW} in this context.)


\pf To show that (A) $\imp$ (B) note that when
  $\Phi$ has the form  (\ref{eq:holv}),
\bee
   (I \ot \Phi)(\Gamma) & = &
    \sum_k R_k ~ T_2 \Big( \sqrt{E_k} \, \Gamma  \sqrt{E_k} \Big) \\
      & = & \sum_k \gamma_k R_k \ot Q_k
\eee
where $T_2$ denotes the partial trace,
$\gamma_k = \tr E_k \Gamma$ and  $Q_k = \ds{
  \frac{1}{\gamma_l} \, T_2 \Big( \sqrt{E_k} \, \Gamma \sqrt{E_k} \Big)}$.
Thus, for arbitrary $\Gamma$, $(I \ot \Phi)(\Gamma)$ is separable.  

The implication (B) $\imp$ (C) is trivial.  To see that (C) $\imp$ (A),
observe that since $(I \ot \Phi)( | \beta \kb \beta | )$  is
separable,  one can find normalized vectors $|v_n \ket$
and $|w_n \ket$ for which
\be
  (I \ot \Phi)( | \beta \kb \beta | ) & \equiv & \frac{1}{d}
    \sum_{jk} | j \kb k | \ot \Phi(| j \kb k |) \\
   & = & \sum_n p_n | v_n \kb v_n | \ot | w_n \kb w_n | \label{eq:Hdeck}
\ee
Now let $\Omega$ be the map
\be
     \Omega(\rho) = d \sum_n | w_n \kb w_n | \,
     \tr \, \Big(\rho \, p_n | v_n \kb v_n | \Big) .
\ee
Then one easily verifies that
\bee
  (I \ot \Omega)( | \beta \kb \beta | ) & = &  \sum_{jkn}
    | j \kb k | \ot | w_n \kb w_n | p_n \bra j , v_n \kb v_n, k \ket \\
  & = & \sum_n p_n |v_n \kb v_n| \ot | w_n \kb w_n |
\eee
where we have used $| v_n \ket = \sum_j |j \kb j , v_n \ket$.
Since a map $\Phi$ is uniquely determined by its action on the
basis $| j \kb k |$, and hence by the action of
$(I \ot \Phi)$ on $ | \beta \kb \beta | $, we can conclude
that $\Phi = \Omega$.  For trace-preserving maps,
we also need to  verify that
$\{d \,  p_n  |v_n \kb v_n| \}$ is a POVM.  Taking the partial trace
of (\ref{eq:Hdeck}), and using the fact that $\Phi$ is trace-preserving
yields
\bee
   T_2 \Big[(I \ot \Phi)( | \beta \kb \beta | ) \Big] & = &
     \frac{1}{d}  \sum_{jk} | j \kb k | \ot  \tr \, ( | j \kb k |) =
  \frac{1}{d} I  \\
    & = &  \sum_n p_n | v_n \kb v_n |
\eee
which is the desired result.   Moreover, we have also shown that
(C) $\imp$ (D).

To show that (D) $\imp$ (A), suppose that
$\Phi(\rho) = \sum_k A_k \rho A_k^{\dg}$ with
$A_k = |w_k \kb u_k |$.  Then the map $\Phi$
can be written in the form (\ref{eq:holv}) with $R_k = |u_k \kb u_k |$.
Moreover, when  $\sum_k A_k^{\dg}  A_k = I$, then
  $\sum_k |u_k \kb u_k | = I$ so that $F_k = |u_k \kb u_k |$
defines a POVM.

The equivalence of (E) and (B) follows easily from the
fact that a density matrix $\Gamma$ is separable if
and only if $(I \ot \Omega)(\Gamma) > 0$ for all
positivity preserving maps $\Omega$ \cite{H3}.  To see that
this is equivalent to (F), it suffices to observe that
$\Omega$ is positivity preserving if and only if its adjoint
$\wh{\Omega}$ is and that
$\wh{\Phi \circ \Upsilon} = \wh{\Upsilon} \circ \wh{\Phi}$,
where the adjoint is taken with respect to the Hilbert Schmidt
inner product so that
$\tr [\wh{\Omega}(A)]^{\dg}  B = \tr A^{\dg} \Omega(B)$.

It may be interesting to recall that $\Upsilon$ is trace-preserving
if and only if  $\wh{\Upsilon}$ is unital so that the adjoint of
a positivity and trace preserving map preserves POVM's.   Thus,
when $\Phi$ has the form (\ref{eq:holv}), the map $\Phi \circ \Upsilon$
is achieved by replacing $E_k$ by $\wh{\Upsilon}(E_k)$.

\medskip

Conditions (E) and (F) could be weakened slightly since it would suffice
to check either for all $ \Upsilon$ in some set of entanglement
witnesses for the space on which $\Phi$ acts.
   However, one does not expect to be able to weaken them beyond this.
Indeed, \cite{DSSTT} and \cite{H.MPR}
contain examples of a channels which preserve 
PPT entanglement, but break other types, i.e., the channel output 
$(I \ot \Phi)(\Gamma)$ is entangled, yet 
the partial transpose $(I \ot T)$ acting on it always yields a positive semi-definite
state $(I \ot T \circ \Phi)(\Gamma) \geq 0$. 

Alternatively, one could also consider maps which are not EBT,
but break particular types of entanglement.



\section{Extreme points}

We now give some results about the extreme points of the convex
set of EBT maps.
In this section we will use some additional results from Choi~\cite{Choi}
who observed that $\Phi$ is completely positive if and only if
$(I \ot \Phi)( | \beta \kb \beta | )$ is positive semi-definite.
When $\Phi$ is written   in the  operator sum form
\be  \label{eq:Kraus}
\Phi(\rho) = \sum_k A_k \rho A_k^{\dg}
\ee
  the Kraus operators $A_k$ can be chosen as the eigenvectors
of $(I \ot \Phi)( | \beta \kb \beta | )$ with strictly
positive (i.e., non-zero)
eigenvalue.  (See Leung \cite{L} for an nice exposition.)
Choi~\cite{Choi} also showed that  $\Phi$ is extreme in the set of CPT
maps if and only if the set $\{ A_j^{\dg} A_k \}$ is linearly independent.
Since both (\ref{eq:Kraus}) and this linear independence are preserved when
$A_i \mapsto \sum_j u_{ij} A_j$, a sufficient condition
for $\Phi$ to be an extreme EBT map is that $\{ A_j^{\dg} A_k \}$
is linearly independent for some set of operators $\{ A_k \}$
satisfying (\ref{eq:Kraus}).
Note that the condition that $\Phi$ is also trace-preserving
becomes $\sum_k A_k^{\dg} A_k = I$.

Recall that an
{\bf extreme CQ map} is one which can be written in the form
\be \label{eq:CQext}
   \Phi(\rho) = \sum_k  |\psi_k \kb \psi_k| \, \bra e_k, \rho \, e_k \, \ket
\ee
with the vectors $\{ e_k \}$ orthonormal.
We can summarize our results as follows.
\begin{thm}  \label{thm:extCQ}  ~~~

\begin{itemize}

\item[A)] If $\Phi$ is an extreme CQ map, then $\Phi$ is an
extreme point in the set of EBT maps.

\item[B)] If $\Phi$ is an extreme CQ map, then $\Phi$ is an
extreme point in the set of CPT maps if and only if
$\bra \psi_j , \psi_k \ket \neq 0 ~~ \forall ~ j,k$
when it is written in the form (\ref{eq:CQext}).

\item[C)] If $\Phi$ is both in the set of EBT maps and an
extreme point of the CPT maps, then $\Phi$ is an extreme
CQ map.


\item[D)]  When $d = 2$, the extreme points of the set of EBT maps
are precisely the extreme CQ maps.  When
$d \geq 3$ there are extreme EBT maps which
are not CQ.

\end{itemize}
\end{thm}

\pf  To prove (A) we assume that
$\Phi = a \Phi_1 + (1-a) \Phi_2$ with $\Phi_1, \Phi_2 \neq \Phi$
$0 < a < 1$ and $\Phi_1, \Phi_2$ both EBT.  Both $\Phi_1, \Phi_2$
can be written in the  form (\ref{eq:rank1}).  By combining these,
one finds one can write
\be \label{eq:qc.cc}
   \Phi(\rho) = \sum_j t_j |\phi_j \kb \phi_j| \, \bra f_j, \rho f_j  \ket
\ee
with $\Phi_1, \Phi_2$ having the same form, but different $t_j \geq 0$.
By assumption, $\Phi$ can be written in the form (\ref{eq:CQext})
with $| e_k \ket$ orthonormal so that
\be  \label{eq:ext.pf}
   \Phi(|e_k \kb e_k|) = |\psi_k \kb \psi_k| =
    \sum_j t_j |\bra e_k, f_j \ket |^2 \, |\phi_j \kb \phi_j|  .
\ee
Since all $t_j \geq 0$, the rank one projection $|\psi_k \kb \psi_k| $
is a linear combination with non-negative coefficients of the
projections  $|\phi_j \kb \phi_j|$.  This is possible only if
those projections  $|\phi_j \kb \phi_j|$ which have  non-zero
coefficients in (\ref{eq:ext.pf})
are identical to the projection $ |\psi_k \kb \psi_k|$.  Hence,
we can conclude that every projection $|\phi_j \kb \phi_j|$
in (\ref{eq:qc.cc}) is equal to one of the projection
$|\psi_k \kb \psi_k| $ in (\ref{eq:CQext}).
Let us now relabel the projections
$|\psi_{k^{\prime}} \kb \psi_{k^{\prime}}| $ so
that they are all distinct and let
$E_{k^{\prime}} = \sum_{i \in k^{\prime}} |e_i \kb e_i |$ where the sum is
taken over those $e_i$ for which the associated projection in
(\ref{eq:CQext}) is  $|\psi_{k^{\prime}} \kb \psi_{k^{\prime}}| $.
Then $\{ E_{k^{\prime}} \}$
gives  a partition of $I$ into mutually orthogonal projections,
i.e,  a von Neumann measurement, and we can write
(dropping the $^{\prime}$s for simplicity)
\be
  \Phi(\rho) = \sum_k  |\psi_k \kb \psi_k| \,
    \tr E_k \rho.
\ee
We can also write
\be
  \Phi_1(\rho) & = & \sum_k  |\psi_k \kb \psi_k| \,
    \tr F_k \rho \\
   \Phi_2(\rho) & = & \sum_k  |\psi_k \kb \psi_k| \,
    \tr G_k \rho
\ee
with $\{ F_k \}$ and $\{ G_k \}$ each a POVM.
Since the $|\psi_{k^{\prime}} \kb \psi_{k^{\prime}}|$ were chosen
to be distinct and the $E_{k^{\prime}}$ orthonormal, it
follows that $\Phi = a \Phi_1 + (1-a) \Phi_2$ if and only if
$E_k = a F_k + (1-a) G_k$.   Since $0 \leq F_k, G_k \leq I$,
this is possible only if $F_k = G_k = E_k$.  But then we
have shown that $\Phi_1 = \Phi_2 = \Phi$, which  proves part (A).

To prove  (B)
note  that the Kraus operators can be chosen as
   $A_k = | \psi_k \kb v_k |$.  Thus,
$A_j^{\dg} A_k = \bra \psi_j , \psi_k \ket | e_k \kb   e_j|$
which yields a linearly independent set if and only if {\em none} of the
$\psi_j$ are mutually orthogonal.  But this is precisely Choi's
condition for  the map to be extreme in the set of all CPT maps.

The proof of part (C) requires Lemma~\ref{lemma:dsep} which is
of interest in its own right.  The proof of (D) when $d = 2$
is given in the following paper \cite{EBQ} on qubit EBT maps,
while the counter-example establishing (D) for $d > 3$ is
given below.

\rmk Recall that
  a QC map can be written in the form
\be
  \label{eq:QCext}
   \Phi(\rho) = \sum_k  |e_k \kb e_k| \, \tr \, \rho F_k
\ee
with the vectors $\{ e_k \}$ orthonormal.  Such maps
can never be extreme in the set of CPT maps; their Kraus
operators always include a subset of the  form
$A_k = |e_k \kb v_k | G_k$ which can {\em not} satisfy
Choi's linear independence condition due to the
orthogonality of the $\{ e_k \}$.  In the case of qubits,
QC maps are not even extreme in EBT, unless they are
also CQ.   However, for $d = 4$, one can have extreme
EBT maps which are QC but not CQ.

\noindent{\bf Example:} Let $\{ g_k \}$ be
orthonormal and consider the POVM consisting of a ``trine''
on  span$\{ g_1, g_2 \}$ and the projection on  span$\{ g_3, g_4 \}$,
i.e.,
  $$E_1 = \ts{\frac{2}{3}}|g_1\kb g_1|,~
  E_2 =\ts{\frac{2}{3}} |g_+\kb g_+|,~ E_3 = \ts{\frac{2}{3}} |g_+\kb g_+|, ~
E_4 = |g_3\kb g_3| + |g_4\kb g_4| $$ where
$| g_{\pm}\ket = \half |g_1 \ket \pm
     \frac{\sqrt{3}}{2} |g_2 \ket $.  Then
$\Phi(\rho) =  \sum_{k = 1}^4   |e_k \kb e_k| \, \tr \, \rho E_k $
is an extreme EBT map, which is QC, but not CQ.

To see that $\Phi$ is extreme it suffices to observe that it
is essentially the direct sum of maps $\Phi_A \oplus \Phi_B$
where $\Phi_A: {\bf C}^2 \mapsto {\bf C}^3$ with
$\Phi_A(\rho) =  \sum_{k = 1}^3  |e_k \kb e_k| \, \tr \, \rho E_k $
and $\Phi_B: {\bf C}^2 \mapsto {\bf C}^1$ with
$\Phi_B(\rho) =  |e_4 \kb e_4|$ for all $\rho$.
  $\Phi_A$  is extreme because it is the adjoint of an
extreme CQ map, and $\Phi_B$ is
the only CPT from map ${\bf C}^2$ to ${\bf C}^1$.
We used the fact that proof of part(A) of Theorem~\ref{thm:extCQ}
extends easily to map from ${\bf C}^d$to  ${\bf C}^{d^{\prime}}$
with $d^{\prime} < d$.

A map which is both CQ and QC projects a density matrix $\rho$
onto its diagonal in a fixed orthonormal basis.  One can generalize
this to CPT maps which take a density matrix to its projection onto
a block-diagonal one.  Such maps  have
  the form $\Phi(\rho) = \sum_k E_k \rho E_k$ where
$  E_k $ are the projections in a von Neumann measurement;
they are not EBT when at least one of the projections has rank $> 1$.
The map in the example above is a generalization of CQ
in the sense that it is the composition of a
block diagonal projection  together with an EBT map, and
thus could be regarded as ``block CQ''.
In a similar spirit, one might regard an extreme CQ map for
which the $\psi_k$ can be split into two mutually orthogonal
subsets as ``block QC''.  With respect to CPT,   maps
which are both  block QC and block CQ
could be considered as generalizations of the quasi-extreme points
introduced in \cite{RSW} for stochastic maps on  ${\bf C}^2$.


We now give some results about the number of Kraus
operators associated with EBT maps.
\begin{thm} \label{thm:d}
If a CPT map $\Phi$ can be written with fewer than $d$ Kraus operators,
then it is {\em not} EBT.
\end{thm}
\pf  This follows  from the fact \cite{Choi} that $\Phi$
can always be written using at most
$r \equiv {\rm rank}[(I \ot \Phi)( | \beta \kb \beta | )$ Kraus
operators.   However, it was shown in \cite{HSTT} that if $r < d$,
then $(I \ot \Phi)  ( | \beta \kb \beta |)  $ is {\em not} separable
and, hence, $\Phi$ does not break the entanglement of the
state $| \beta \kb \beta |$.  Alternatively, one
could observe that if $r < d$, then
at least one eigenvalue of $(I \ot \Phi)( | \beta \kb \beta |$
is greater than $1/d$, while its left reduced density matrix
has all eigenvalues equal to $1\over d$ (since $\Phi$ is CPT).
However, in Ref. \cite{H.MP} it was shown that if  a state is separable,
then its the maximal eigenvalue must not exceed the maximal eigenvalue of
either of subsystems.    ~~~\qed

\begin{lemma}
If $\Phi$ is a CPT map for which
  ${\rm rank}[(I \ot \Phi)( | \beta \kb \beta | )] = d$, then
$\Phi$ is EBT if and only if $T \circ \Phi$
is completely positive.
\end{lemma}
This follows immediately from a (non-trivial) result in
\cite{HLVC} which implies that a $d^2 \times d^2$ density
matrix of rank $d$ is separable if and only if it has
positive partial transpose.

The following lemma is of some interest since
  one can find examples  \cite{DTT} of separable matrices of rank $d$
whose decomposition into product pure states requires more
than $d$ products.  The additional hypothesis that the reduced
density matrix $\rho_A  = \tr_B \rho$ also has rank $d$ is
crucial. The lemma was first proven in \cite{HLVC}. Here we present
a simpler proof.

\begin{lemma}  \label{lemma:dsep}
Let $\rho$ be a density matrix on
${\cal H}_A \ot {\cal H}_B$.  If $\rho$ is separable,
$\rho$ has rank $d$, and
$\rho_A  = \tr_B \rho$ has rank $d$, then $\rho$ can
be written as a convex combination of products of pure
states using at most $d$ products.
\end{lemma}
\pf Since $\rho$ is separable it can be written in the form
\be   \label{eq:minprod}
   \rho = \sum_{i=1}^k \lambda_i \, \proj{a_i} \otimes \proj{b_i}.
\ee
Assume that $k > d$ and that $\rho$ can not be written in
the form (\ref{eq:minprod}) using less than $k$ products.
Since $\rho_A$ has exactly rank $d$, there is no loss of
generality in assuming that the vectors above
have been chosen so that $|a_1 \ket, |a_2 \ket, \ldots |a_d \ket$
are linearly independent.   Moreover,
since $\rho$ has rank $d<k$,
the first $d+1$ vectors $|a_i \ket \otimes |b_i\ket $
must be linearly dependent so that one can find $\alpha_j$  such that
\be
\sum_{j=1}^{d+1} \alpha_j \, |a_j \ket \ot  |b_j\ket  = 0.
\ee
Now let $\{ |e_k \ket \}$ be an orthonormal basis for ${\cal H}_B$.
Then
\be   \label{eq:ldA}
\sum_{j=1}^{d+1} \alpha_j \bra e_k, b_j \ket \, |a_j \ket   = 0 ~~~\forall ~k.
\ee
Since the first $d$ vectors $|a_j \ket $ are linearly
independent, there is a vector ${\bf x}$ in ${\bf C}^{d+1}$
such that $\sum_j v_j |a_j \ket = 0$ if and only if
${\bf v}$ is a multiple of ${\bf x}$.
Applying this to the coefficients in (\ref{eq:ldA})
one finds that there are numbers
$\nu_k$ such that $u_j \bra e_k, |b_j \ket = \nu_k x_j$.
Let $|\nu \ket$ be the vector $\sum_k \nu_k |e_k \ket$.
Then $\alpha_j |b_j \ket = x_j |\nu \ket$.  Since $|b_j \ket$
was chosen to have norm $1$, it follows that when $\alpha_j \neq 0$,
$\left|\frac{x_j}{\alpha_j} \right| = 1$ and $|b_j \ket = e^{i 
\theta_j} |\nu \ket$.
Thus, one can rewrite (\ref{eq:minprod}) as
\be  \label{eq:alt.rho}
   \rho = \sum_{j : \alpha_j \neq 0} \lambda_j \,
        \proj{a_j} \otimes \proj{b_j}  + \sum_{j : \alpha_j = 0}
    \lambda_j  \, \proj{a_j} \otimes \proj{\nu} .
\ee
Suppose that $t$ of the $\alpha_j$ are non-zero.
Since the vectors $\{ a_j \ket : \alpha_j \neq 0 \}$ are
linearly dependent, the density matrix
$\wt{\rho}_A = \sum_{j : \alpha_j = 0}  \lambda_j  \, \proj{a_j}$
has rank strictly $< t$ and can be rewritten in the form
$\wt{\rho}_A = \sum_{k = 1}^{t'} \lambda_j' |a_j' \kb a_j' |$
using only $t' < t$ vectors.  Substituting this in (\ref{eq:alt.rho})
gives $\rho$ as linear combination of products using strictly
less than $k$ contradicting the assumption that (\ref{eq:minprod})
used the minimum number.

\smallskip


\noindent{\bf Proof of (C):}
   If $\Phi$ can be written with fewer than $d$ Kraus operators,
it is not entanglement breaking; and if it requires more than $d$
Kraus operators, it is not extreme.   Hence we can assume that
${\rm rank}[(I \ot \Phi)( | \beta \kb \beta | ) = d$.
The result then follows from Lemma~\ref{lemma:dsep}.

\smallskip

We now show that, for $d = 3$, the set of entanglement breaking maps
has extreme points which are not CQ.  Moreover, unlike the $d = 4$
example considered earlier, there is no decomposition into
orthogonal blocks associated with this map.

\noindent{\bf Counterexample:}
Let $|0\ket, |1\ket, |2\ket$ be an orthonormal basis for ${\bf C}^3$
and consider the following four vectors corresponding to the vertices
   of a tetrahedron
    \begin{eqnarray*}
   |v_0 \ket & = & \frac{1}{\sqrt{3}}\Big(+ |0\ket + |1\ket + |2\ket \Big)   \\
    |v_1 \ket  & = & \frac{1}{\sqrt{3}}
      \Big(+ |0\ket - |1\ket - |2\ket \Big)  \\
     |v_2 \ket   & = & \frac{1}{\sqrt{3}}
   \Big(-|0\ket + |1\ket - |2\ket \Big)    \\
     | v_3 \ket & = & \frac{1}{\sqrt{3}}\Big(-|0\ket - |1\ket + |2\ket \Big)
   \end{eqnarray*}
and let
\be
   \Phi(\rho) = \frac{3}{4} \sum_{i=0}^3 |v_i \kb v_i| \, \tr \rho |v_i \kb v_i|
\ee
We now show that $\Phi$ is an extreme point for the set of
   entanglement-breaking maps.  To see this, first recall
   that any entanglement breaking map $\Psi$ can be written as
\be
   \Psi(\rho) = \sum_{i} \alpha_i |y_i\kb y_i | \tr \rho |z_i\kb z_i |
\ee
   Let $\Psi$ be one of the entanglement breaking maps whose
   convex combination is $\Phi$, and
   let $| y \ket$ and $|z \ket$ be $|y_i\ket $ and $|z_i\ket$
   for some fixed $i$ in this above expression for $\Psi$.
   Now, consider the six vectors $|w_{ij}\ket$
   for $i < j$,
   where these are defined so that $\braket{w_{ij}}{v_k} = 0$ for $k\neq
i,j$.   For example, $|w_{01}\ket = \frac{1}{\sqrt{2}}( |1 \ket + |2
\ket)$.
   Then,
\be
   \Phi(|w_{ij} \kb w_{ij} |) = \frac{1}{2}\left(|v_i \kb v_i| +
    |v_j \kb v_j| \right)
\ee
   so for input $|w_{ij} \kb w_{ij} |$, the output
   has rank 2 and is orthogonal to $w_{kl}$,
   where $i,j,k,l$ are all distinct.
   We thus have that for $|y \ket$ and $|z \ket$,
\be
   \braket{w_{ij}}{y} = 0 \mathrm{\ \ \ or \ \ \ }
   \braket{w_{kl}}{z} = 0
\ee
   where $\{i,j,k,l\}$ is any permutation of $\{0,1,2,3\}$, as above.

   Now, consider $| y \ket$.
  Suppose it is orthogonal to two of $w_{01}$, $w_{02}$, and $w_{12}$.
   Then, we must have $| y \ket =  |  v_3 \ket$.
   This means that $| y \ket$ is not orthogonal to $w_{23}$, $w_{13}$ and
   $w_{03}$, which implies in turn
   that $|z \ket$ is orthogonal to $w_{01}$, $w_{02}$
   and $w_{12}$, showing that $|z \ket = |  v_3 \ket$ as well.

   The other case is when $y$ is not orthogonal to at least
   two of the above three vectors $w_{01}$, $w_{02}$, $w_{12}$;
   we can assume by symmetry that these two are $w_{01}$ and
   $w_{02}$.  Then $z$ is orthogonal to $w_{23}$ and $w_{13}$, showing that
   $|z \ket = |  v_0 \ket$.  By the same reasoning as in the
   last paragraph, we now have that $| y \ket = |  v_0 \ket$ as well.

   Thus, all the $y_i$ and $z_i$ in the above expression for $\Psi$ must
   be one of the four vectors $v_j$.  It follows easily from this that
   $\Psi = \Phi$.   Moreover, we have shown that the Holevo form
for $\Phi$ is essentially unique.   Hence $\Phi$ can not
be written in the form required for it to be a CQ map.  ~~\qed

   Note that  $\Phi$ is not extreme in the set of CPT maps.
In fact, it  can be represented as a convex combination of CPT maps
in several ways.  For example, it can be written as
    the convex combination of the identity map,
   with weight $\frac{1}{3}$,
   and the average of the three CP
  maps that first project the state into one of the three
   planes $\{|0\ket, |1\ket  \}$, $\{|0\ket,  |2\ket \}$,
   $\{|1\ket, |2\ket \}$, and then apply the $\sigma_x$ operator
   for that plane  interchanging the two basis states, with weight
   $\frac{2}{3}$.  It can also be written as a convex
combination of the identity and the four maps
corresponding to conjugation with a unitary map which reflects
across the plane orthogonal to one of the vectors $|v_j \ket $.


\section{Representations in bases}

  Let $G_0 = d^{-1/2} I$ and let $G_1 \ldots  G_{d^2 -1}$
be a basis for the subspace of self-adjoint $d \times
d$ matrices with trace zero
which is  orthonormal in the sense $\tr G_j^* G_k = \delta_{jk}$.
Then $\{ G_k \}$, $k = 0, 1  \ldots d^2 \! - \! 1$ is an orthonormal basis
for the subspace of self-adjoint $d \times d$ matrices and every
density matrix can be written in the form
\be
\rho = \frac{1}{d}I + \sum_{j=1}^{d^2-1} w_j G_j  =
    \sum_{j=0}^{d^2-1} w_j G_j
\ee
with  $w_j = \tr \rho G_j$ so that  $w_0 = d^{1/2}$.  It then follows that
\bee
    \sum_{j=0}^{d^2-1} w_j^2 = \tr \rho^2 \leq \tr \rho = 1 ~~{\rm and}~~
     \sum_{j=1}^{d^2-1} w_j^2  \leq \frac{d-1}{d}  .
\eee
Then any linear (and hence stochastic) map $\Phi$ on the
self-adjoint $d \times d$ matrices can be represented as a
$d^2 \times d^2$ matrix $\bT$ with elements
$t_{jk} = \tr G_j \Phi(G_k)$.
Now let $\Phi$ be a Holevo channel
with density matrices $R_k = \sum_j w_j^k G_j$ and POVM
$F_k = \sum_n u_n^k G_n$ ($k = 1 \ldots N$)
and write $\rho = \sum_i x_i G_i$.
Then it is straightforward to verify that
$t_{jn} = \sum_k w_j^k u_n^k$.  Thus,   $\bT = W^T U$
where $W$ and $U$ are the $d^2 \times N$ matrices with elements
$w_{jk} = w_j^k$ and $u_{nk} = u_n^k$ respectively.  The condition
that $\{ F_k \}$ is a POVM is precisely that the first row of
$\bT$ is $(1, 0, \ldots , 0)$.

Such representations have been studied in more detail for qubits
using the Pauli matrices for $G_k$.   Recently, several
generalizations have been consider for $d = 3$
\cite{King3} and higher \cite{Cortese,PR}.   Another natural choice of
basis  has $G_{jk} = |j \kb k|$ for some orthonormal
basis $| j \ket$.   In this case some modifications are
needed since $I = \sum_k G_{kk}$.  For $j < k$, one could also
replace $G_{jk},  G_{kj}$ by $2^{-1/2}(G_{jk} \pm G_{kj})$
which act like $\sigma_x$ and $i \sigma_y$
for the two-dimensional subspace span$\{ |j\ket, |k \ket \}$.
Unfortunately, when $d > 2$, the requirement that $R_k$ and $F_k$ are
positive semi-definite does not seem easily
related to a condition between $u_0$ and $\sum_{j=1}^{d^2-1} u_j^2$
in any of these bases.  Hence, such representations
  seem  most useful for qubits, as discussed in \cite{EBQ}.

For a CQ or QC channel, $W$ and $U$ are $d^2 \times d$ which
implies rank($\bT$) $\leq d$.   Hence the image of a QC
or CQ channel lies in a subspace of dim $\leq d-1$.
This raises the question of whether or not a stochastic map
for which the image of the set of density matrices lies in
a subspace of sufficiently small dimension is always entanglement
breaking.
(This is true  for   qubits for which all planar maps are EBT.)

  For a basis in which a necessary condition for positive
semi-definiteness is $\sum_{i = 1}^{d^2-1} |x_i|^2 \leq x_0^2$,
one can show that EBT implies
$\sum_{j = 1}^{d^2-1} |t_{jj}| \leq 1$.
  For details, see  Ref. \cite{EBQ}

In general, a matrix $\bT$ can be written as a product in many
ways.   We have shown that $\bT$ represents an entanglement-breaking
map if it can be decomposed into a product $\bT = W^T U$ whose
elements $W, U$ have very special properties.   There is also
a correspondence between  the matrix $\bT$ which represents
$\Phi$ in a basis in the usual sense and the matrix
$(I \ot \Phi)( | \beta \kb \beta | )$.  It would seem that
the requirement that $(I \ot \Phi)( | \beta \kb \beta | )$
is separable is related to the product decomposition of $\bT$;
however, we have not  analyzed this.
It may be more amenable to the filtering approach advocated
by Verstraete and Verschelde \cite{VV}.

\bigskip


\noindent{\bf Acknowledgment:} Part of this work was done while the
authors participated  in the program on Quantum Computation
at the Mathematical Sciences Research Institute at Berkeley
in November, 2002.

The work of M.H. is supported by EC, grant EQUIP (IST-1999-11053),
RESQ (IST-2001-37559) and QUPRODIS (IST-2001-38877).
The work of M.B.R, was partially supported  by
  the National Security Agency (NSA) and
  Advanced Research and Development Activity (ARDA) under
Army Research Office (ARO) contract numbers
    DAAG55-98-1-0374 and DAAD19-02-1-0065, and by the National Science
         Foundation under Grant number DMS-0074566.

\bigskip


\end{document}